\documentclass[twocolumn,longbib,trackchanges]{aastex701}

\usepackage{amsmath}
\usepackage{amssymb}
\usepackage{mathtools} 
\usepackage{graphicx} 
\usepackage[capitalize]{cleveref}
\usepackage{lineno}

\makeatletter
\g@addto@macro\bfseries{\boldmath}
\makeatother

\newcommand{\dd}{\mathrm{d}}

\newcommand{\eos}{EoS}
\newcommand{\eoss}{EoSs}

\newcommand{\bl}{\beta_\mathrm{L}}
\newcommand{\pl}{p_\mathrm{L}}
\newcommand{\nl}{n_\mathrm{L}}
\newcommand{\mul}{\mu_\mathrm{L}}
\newcommand{\muc}{\mu_\mathrm{c}}

\newcommand{\el}{\varepsilon_\mathrm{L}}
\newcommand{\bh}{\beta_\mathrm{H}}
\newcommand{\ph}{p_\mathrm{H}}
\newcommand{\nh}{n_\mathrm{H}}
\newcommand{\muh}{\mu_\mathrm{H}}
\newcommand{\eh}{\varepsilon_\mathrm{H}}

\newcommand{\vlh}{V_{\bl,\bh}}

\newcommand{\Ulh}{\mathcal{U}_{\bl,\bh}(\beta)}

\begin{document}

\title{Constrained Gaussian-process bridge prior for neutron-star equation-of-state inference}

\date{\today}

\author[0000-0003-3469-7574]{Tyler Gorda}
\affiliation{Center for Cosmology and AstroParticle Physics (CCAPP), Ohio State University, Columbus, OH 43210}
\affiliation{Department of Physics, The Ohio State University, Columbus, OH 43210, USA}
\email{gorda.1@osu.edu}

\author[0000-0002-2188-3549]{Oleg Komoltsev}
\affiliation{Institut für Theoretische Physik, Goethe Universität,
Max-von-Laue-Str. 1, 60438 Frankfurt am Main, Germany}
\email{komoltsev@itp.uni-frankfurt.de}

\author[0000-0001-7991-3096]{Aleksi Kurkela}
\affiliation{Faculty of Science and Technology, University of Stavanger, 4036 Stavanger, Norway}
\email{aleksi.kurkela@uis.no}

\author{Eirik Sunde}
\affiliation{Faculty of Science and Technology, University of Stavanger, 4036 Stavanger, Norway}
\email{ei.sunde@stud.uis.no}

\begin{abstract}
    We set forth a new method for generating model-agnostic, nonparametric priors for neutron star equation-of-state inference that are stable, causal and thermodynamically consistent by construction.
    This generalizes Gaussian processes to include global thermodynamic constraints, specifically allowing the inclusion of any number of training points in the form $(\mu, n, p)$ while retaining thermodynamic consistency between them.
    The method is based on constructing constrained Gaussian-process bridges, whose correlation properties can be tuned at will allowing flexibility between a conservative prior and a theory-informed prior.
    The method does not require any shooting to obey multiple constraints and provides an efficient and informed way to include both chiral effective field theory and perturbative quantum chromodynamics constraints within the same framework.
\end{abstract}


\section{Introduction}
\label{sec:intro}

Progress in astrophysical observations of neutron stars (NSs) has opened a pathway into studying the densest matter known to exist, found deep within their cores \citep{Demorest:2010bx,Antoniadis:2013pzd,TheLIGOScientific:2017qsa,Riley:2019yda,Miller:2019cac}. These cores can reach densities up to seven times the nuclear saturation density, $n_s \approx 0.16\,\mathrm{fm}^{-3}$, corresponding to values exceeding twice the density inside of a neutron. 
Understanding the phase structure and properties of baryonic matter under such extreme conditions is of great interest \citep{Annala:2019puf}. 

Inferring the microphysical conditions within neutron-star cores from macroscopic observables relies on equation-of-state (\eos) inference, which connects the thermodynamic properties of dense matter to the global properties of neutron stars. 
Each candidate \eos\ predicts characteristic relations among stellar masses, radii, and tidal deformabilities, as well as the gravitational-wave spectra produced during binary mergers \citep{Oppenheimer:1939ne, Tolman:1939jz,Hinderer:2009ca,Baiotti:2016qnr,Ecker:2024uqv}. 
Comparing these theoretical predictions with astrophysical observations enables one to rule out microphysical models that are inconsistent with current neutron-star data.

Two main strategies have been pursued to exploit the growing body of neutron-star observations. 
In the first approach, the \eos\ is derived from microphysical models of dense matter, constructed under specific theoretical assumptions and approximations. 
These models, or their parameter ranges, can then be constrained or excluded by observational data \citep{Oertel:2016bki,Baym:2017whm, Cartaxo:2025jpi}.

With the increasing abundance and precision of neutron star observations, a complementary model-agnostic strategy for \eos\ inference has become widely adopted \citep{Hebeler:2013nza,Kurkela:2014vha,Steiner:2015aea,PhysRevLett.120.172703,PhysRevC.98.045804,LIGOScientific:2018hze,Landry:2018prl,Annala:2019puf,Capano2020,Landry:2020vaw,Essick:2020flb,Dietrich:2020efo,PhysRevLett.126.061101,Raaijmakers_2021,Miller:2021qha,Huth2022,Lim:2022fap,PhysRevD.110.034035,Koehn:2024set}.
In this framework, a large ensemble of trial \eoss\ is generated using flexible mathematical representations—such as piecewise polytropes, piecewise speed-of-sound parameterizations, spectral expansions, or Gaussian processes—and subsequently compared against observational data. 
Those inconsistent with the measurements are ruled out, typically through a Bayesian statistical framework, leaving a model-agnostic estimate of the allowed \eos\ parameter space. 
Although this approach does not assume any specific microscopic model, the inferred features of the \eos\ can nonetheless provide valuable insights into the underlying microphysics \citep{Annala:2019puf,Annala:2023cwx}, including possible phase transitions in neutron-star cores \citep{Gorda:2022lsk,Mroczek:2023zxo,Komoltsev:2024lcr,Essick:2023fso}.

Although model-agnostic inference does not assume a specific microscopic model, it is not fully model-independent and the choice of prior strongly influences the posterior, especially at densities where data is scarce.
Priors encode essential physical assumptions—namely that the \eos\ must remain thermodynamically consistent, mechanically stable, and causal, with the speed of sound below the speed of light at all densities.
They can also incorporate theoretical input where first-principles calculations provide controlled constraints. 
Two main strategies are commonly employed. In the first, the prior is anchored to nuclear-theory calculations at densities up to nuclear saturation and extrapolated to higher densities \citep{Hebeler:2013nza}. 
In the second, high-density constraints from perturbative quantum chromodynamics (pQCD) are additionally imposed, and the \eos\ is interpolated between the nuclear and pQCD regimes around 20-40 $n_s$ to obtain a family of models that is consistent at all densities \citep{Kurkela:2014vha}.

The extent to which pQCD inputs influence \eos\ inference at neutron-star densities has been widely discussed, and their qualitative effect of softening the \eos\ at the highest densities is well established \citep{Gorda:2022jvk,Somasundaram:2022ztm, Komoltsev:2023zor, Altiparmak:2022bke, Koehn:2024set, Finch:2025bao, Komoltsev:2025vwn}. 
Implementing priors that simultaneously enforce thermodynamic consistency, causality, and stability while incorporating pQCD constraints poses significant technical challenges, as these requirements strongly restrict the functional space of admissible \eoss. 
Such constraints can be satisfied within parametric priors—such as piecewise polytropes \citep{Kurkela:2014vha} or piecewise speed-of-sound interpolations \citep{Annala:2019puf}—however, it has been shown that parametric representations may act as biased estimators of the \eos\ leading to uncontrolled correlations among inferred quantities \citep{Legred:2022pyp}. 
Nonparametric Gaussian-process (GP) priors alleviate some of these issues but make it difficult to impose all physical constraints simultaneously.

\begin{figure*}[t!]
    \centering
    \includegraphics[width=0.9\linewidth]{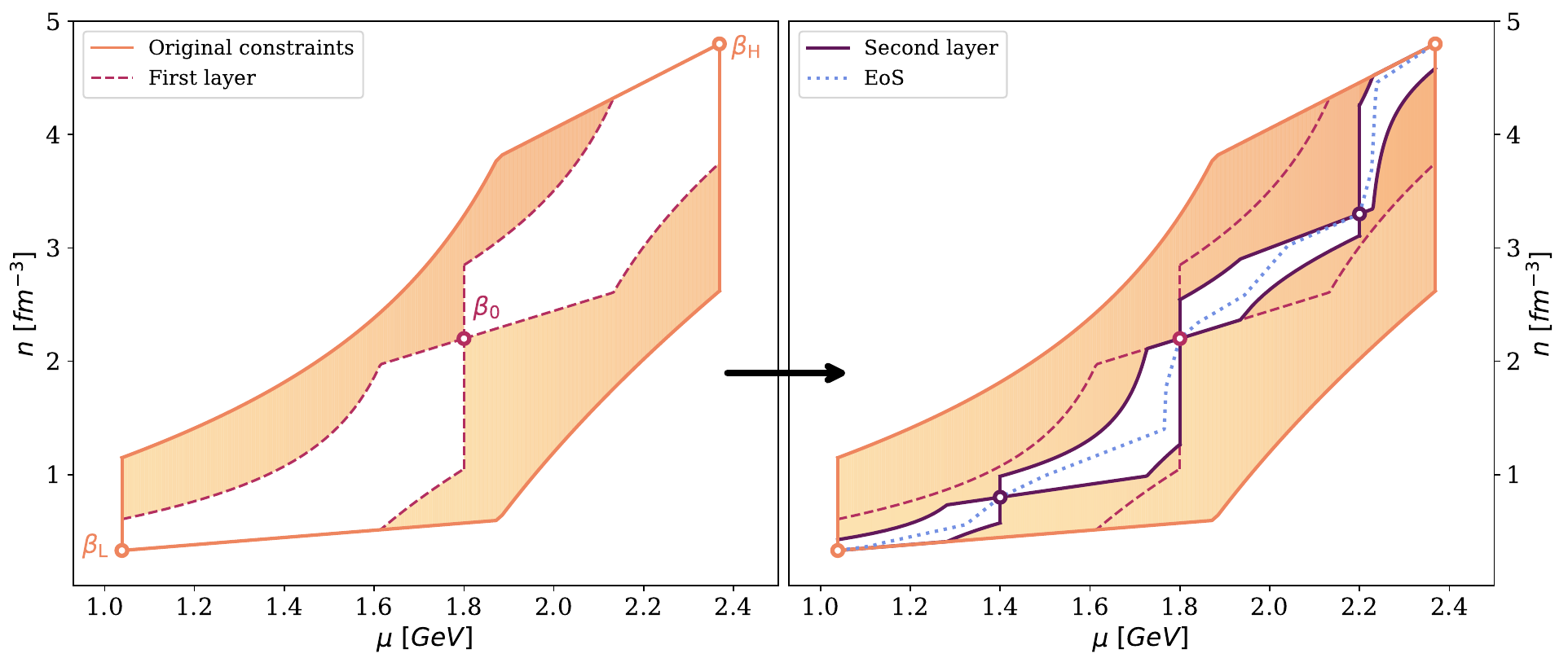}
\caption{Example of a two-step self-similar refinement (fractal). Knowledge of the \eos\ at low and high densities
($\bl = (\mul, \nl, \pl)$ and $\bh = (\muh, \nh, \ph)$)
implies constraints on points $\beta_0$ lying between them.
Once a point $\beta_0$ is chosen, similar constraints are imposed on the
intervals $[\beta_\mathrm{L}, \beta_0]$ and $[\beta_0, \beta_\mathrm{H}]$.
Iteratively refining each interval by introducing new points eventually
leads to a complete \eos\ illustrated by the blue dotted line obtained after
multiple iterations passing through the specified points.}
    \label{fig:fractal_cartoon}
\end{figure*}

Different strategies have been proposed to address these challenges. 
In~\cite{Gorda:2022jvk}, the GP prior was extended up to an intermediate demarcation density above the range realized in stable neutron stars, while the pQCD input at higher densities was incorporated through a maximally conservative criterion that excludes any \eos\ inconsistent with a mechanically stable, causal and thermodynamically consistent connection to the pQCD \eos\ \citep{Komoltsev:2021jzg}.
This treatment effectively separates the \eos\ behavior below and above the demarcation density, introducing dependence on this matching point and allowing very stiff or soft behavior at higher densities, without explicitly inferring the \eos\ beyond the demarcation region. 
In an alternative approach \citep{Finch:2025bao}, the GP prior is simply extended to pQCD densities, removing the need for an intermediate likelihood function; however, because the pQCD constraints are not imposed directly on the prior, most generated \eoss\ receive very small posterior weights and are rejected. 
In another alternative approach using Bayesian Model mixing \citep{Semposki:2025etb}, chiral-effective-field-theory (EFT) and pQCD information is used within their respective regime of validity to condition a GP of the pressure divided by the noninteracting pressure as a function of density. 
This technique allows one to incorporate both the low- and high-density information at the same time, but it does not by construction impose causality at intermediate densities.

In this work, we propose a new method to generate the prior by constructing a constrained GP-bridge-type process between low- and high-density constraints, addressing the current shortcomings of non-parametric interpolation.

An ordinary GP bridge \citep{10.1007/978-3-540-70847-6_15} with a Gaussian covariance structure—corresponding to a radial basis function (RBF) kernel—can be constructed in two steps. 
In the first step, the speed of sound between the two end points is drawn from uncorrelated white noise. 
In the second step, the RBF covariance structure is imposed through deterministic smoothing governed by an appropriate diffusion equation, producing continuous interpolation functions whose correlation lengths are determined by the diffusion parameters. 
Since the diffusion induces only pairwise correlations, this linear smoothing procedure results in a GP with an RBF kernel.

However, such an approach does not naturally satisfy the thermodynamic consistency conditions of the \eos. 
To incorporate these constraints, we propose an alternative procedure. 
Instead of starting from white noise, we generate a space of maximally noisy \eoss\ through a self-similar (fractal) refinement of the \eos, producing structure across all density scales within the allowed functional space formulated in~\cite{Komoltsev:2021jzg}. 
This construction yields \eoss\ with minimal short-range correlations, but with the non-Gaussian long-range correlations required to satisfy global physical constraints. 
The desired short-range RBF covariance structure can then be introduced by applying a diffusion equation, leading to a modified  GP-bridge–type construction that preserves the essential thermodynamic properties of the \eos\ while allowing the correlation length to be tuned as needed.

\section{Method}

In this work, we restrict our discussion to zero-temperature matter in $\beta$-equilibrium. The method consists of two main steps. 
First, we generate a noisy, self-similar (fractal) \eos\ by sampling the full function space allowed by physical constraints. 
In the second step, local correlations are imposed by applying a diffusion equation in the \eos\ space.

\subsection{Robust equation-of-state bounds}

To construct the prior, we first sample the function space connecting the low- and high-density limits of the \eos. 
Each limit is characterized by a triplet of thermodynamic quantities: the chemical potential, number density, and pressure,
\begin{align}
\bl &\coloneq (\mul, \nl, \pl), \\
\bh &\coloneq (\muh, \nh, \ph).
\end{align}
The function space consists of all equations of state represented by $n(\mu)$ that connect these two limits while remaining mechanically stable, causal, and thermodynamically consistent.

\begin{figure*}
    \centering
    \includegraphics[trim=1.cm 7cm 1.cm 1.5cm, clip,width=\linewidth]{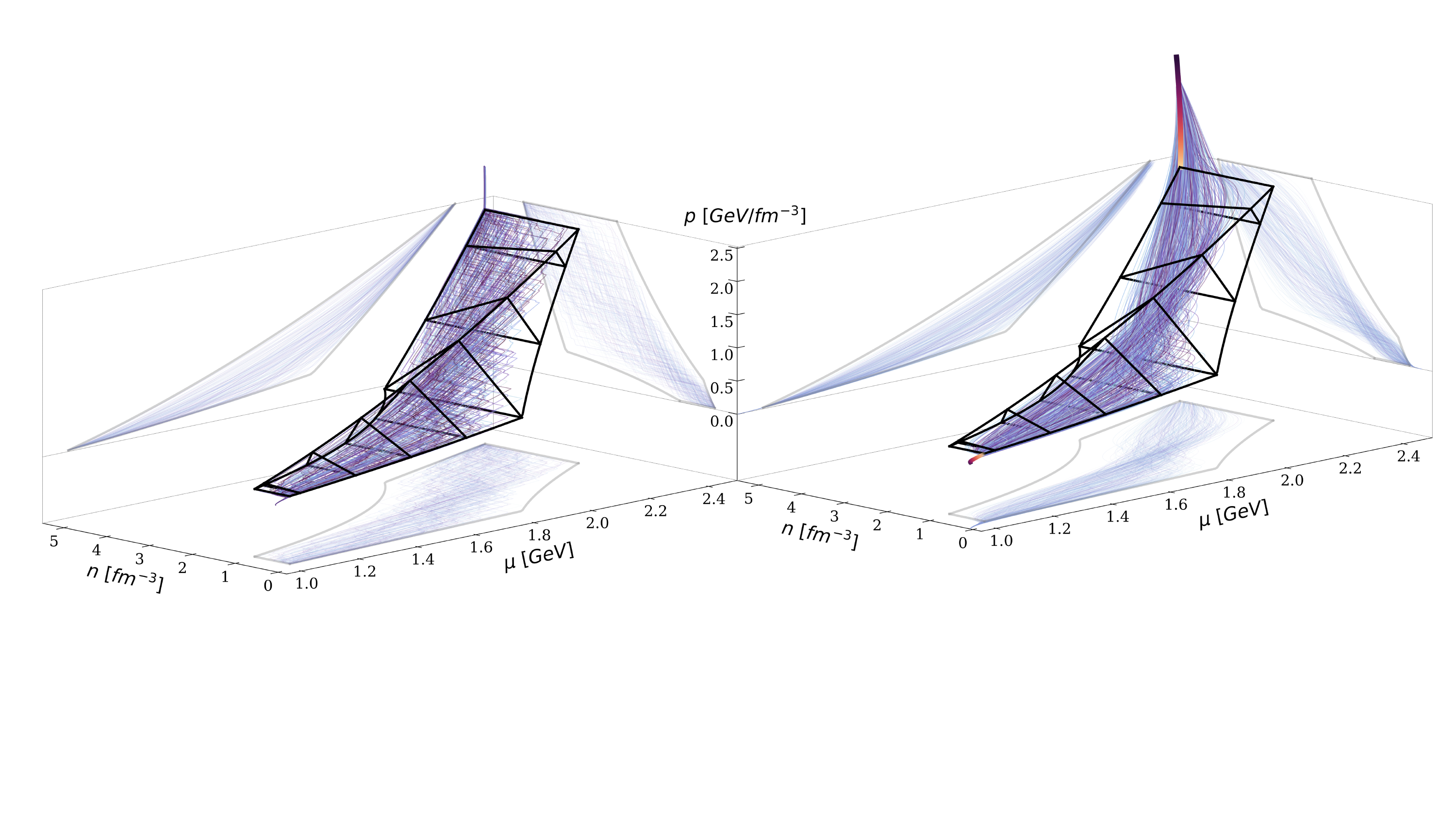}
\caption{Three-dimensional rendering of the volume in the ($\mu, n, p$) that can be reached by a stable, causal and consistent \eos\ interpolating between known low- and high-density limits arising from chiral EFT and perturbative QCD. 
The limits are depicted as the thick lines emerging from the lower-right and upper-left corners of the volume. 
(Left) Sample of fractal \eoss\ constructed using 10 iterations of the self-similar refinement process discussed in Sec.~\ref{sec:fractal}. The coloring of the \eoss\ is arbitrary and used solely to improve visual clarity. 
(Right) Sample of GP-bridge \eoss\ constructed from the self-similar \eoss\ by diffusive smoothing as described in Sec.~\ref{sec:implementation}.}
    \label{fig:3D}
\end{figure*}

Mechanical stability requires $n(\mu)$ to be a single-valued function. Causality demands that the function be monotonic with a sufficiently large derivative,
\begin{align}
\frac{\mu}{n}\frac{d n}{d\mu} \geq 1,
\end{align}
which ensures that the speed of sound does not exceed the speed of light\footnote{For non-differentiable functions, the constraint generalizes to excluding functions for which $\mu n(\mu) \leq \mu' n(\mu')$ for any pair of chemical potentials with $\mu< \mu'$.}.
Thermodynamic consistency requires that the pressure difference between the two limits is correctly reproduced,
$$
\ph-\pl = \int_{\nl}^{\nh} dn\, \mu(n).
$$
In \cite{Komoltsev:2021jzg}, it was shown that, given $\bl$ and $\bh$, any valid intermediate $\beta$ must lie within a volume $\vlh$ (see Fig.~\ref{fig:3D}), defined as follows. For each chemical potential $\mu$, the number density must satisfy $n_{\rm min}(\mu) < n < n_{\rm max}(\mu)$. Furthermore, for each $\mu$ and $n$, the pressure is constrained by $p_{\rm min}(\mu) < p < p_{\rm max}(\mu, n)$, where all relevant equations are given in App.~\ref{sec:3d_volume_eqs}.

\subsection{Self-similar refining:}
\label{sec:fractal}

We generate samples of the \eos\ function space through a process of self-similar refinement.  
Starting from the interval $[\bl, \bh]$, we introduce a new intermediate point $\beta_0$ through which the \eos\ must pass.  
As long as this point lies within the allowed volume $\beta_0 \in \vlh$, there exists at least one function that satisfies all physical constraints and passes through it.

The interval $[\bl, \bh]$ is then divided into two subintervals, $[\bl, \beta_0]$ and $[\beta_0, \bh]$.  
Each subinterval can be further refined by selecting new points from the corresponding subvolumes $\mathcal{V}_{\mathrm{L}0}$ and $\mathcal{V}_{0\mathrm{H}}$.  
Iterating this procedure recursively generates \eoss\ that span the entire space of allowed equations of state while developing structure on all density scales. See Fig.~\ref{fig:fractal_cartoon} for an illustration of this refinement process.

While the geometry of the function space is dictated by physical constraints, defining a prior over this space requires specifying a probability measure that determines how new points are sampled within the allowed volumes $\mathcal{V}_{\beta,\beta'}$.  
This choice corresponds to specifying the prior distribution.  
In the present work, we select intermediate points from a uniform distribution $\mathcal{U}_{\beta,\beta'}$ within each allowed volume, using a Euclidean metric in $(\mu, n, p)$-space.  
Other sampling prescriptions could naturally be adopted, leading to different implicit priors over the \eos\ space. The details how we perform the uniform sampling are relegated to App.~\ref{sec:sampling_method}.

Fig.~\ref{fig:3D} (left) illustrates a sample of such equations of state constructed using 10 layers in the three-dimensional ($\mu, n, p$) - space, together with their projections onto the corresponding two-dimensional planes at fixed values of $\bl$ and $\bh$. The values of $\bl$ and $\bh$ in this figure are illustrative of the values arising from chiral EFT and pQCD at their domains of applicability.

\subsection{Introducing local Gaussian correlations}
\begin{figure*}
    \centering
    \includegraphics[width=1.0\linewidth]
    {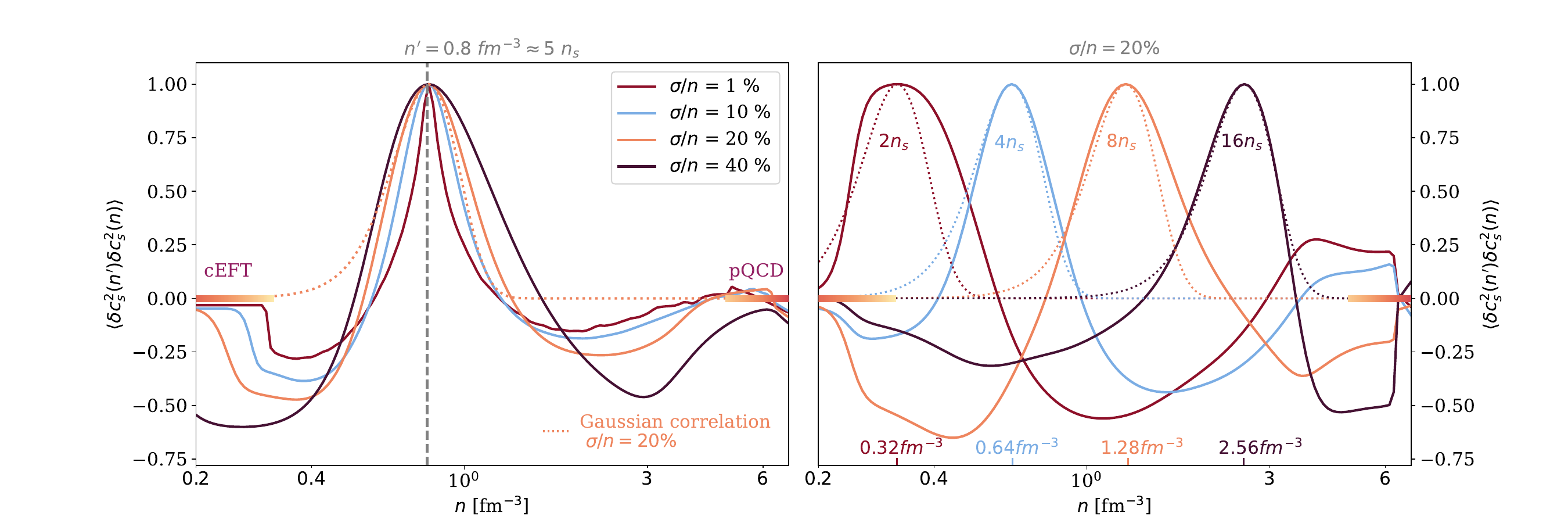}
\caption{
The two-point correlation function of the sound speed, $\langle \delta c_s^2(n')\, \delta c_s^2(n) \rangle$, 
showing correlations between values at $n' = 0.8\ \mathrm{fm}^{-3}$ and different $n$ 
for varying levels of diffusion (left panel), and for a fixed amount of diffusion 
with varying $n'$ (right panel). 
Thin dashed lines show the corresponding Gaussian correlations, 
with correlation lengths set by the amount of diffusion, $\sigma = \sqrt{4\,D(n')\,t}$. 
The local correlations closely follow the Gaussian form, 
but the global structure imposed by physical constraints produces 
a long-range anti-correlation. 
}
    \label{fig:correlator}
\end{figure*}

The above discussed method of generating self-similar refinement processes leads to \eoss\ $\mu(n)$ that have structures at all scales. 
For both numerical and physical reasons, we wish to be able to impose a nonzero correlation length to these functions. 
We do so by applying a heat-kernel smoothing to the resulting processes to produce the desired covariance structure and correlation length.

Specifically, we find it advantageous to diffuse the chemical potential as function of $n$, $\mu(n)$.  
We elevate the equation of state to be a function of an unphysical flow time $\tau$, $\mu(n, \tau)$. The evolution in the flow time is then determined by the heat equation
\begin{equation}
    \partial_\tau \mu(n , \tau) = \partial_n \left[ D(n) \partial_n \mu(n, \tau) \right],
\end{equation}
with the initial condition $\mu(n , 0) = \mu(n)$ and a density-dependent diffusion coefficient $D(n)$ and imposing Dirichlet boundary conditions. 
This procedure increases the correlation between points which are within the diffusion scale $\delta n \sim \sqrt{ 4 D \tau} $ from each other.

It is crucial that the diffusion process preserves the \eos\ within the physically allowed function space.  
The condition of mechanical stability requires $\mu(n)$ to be a monotonically increasing function.  
Since the initial configuration is monotonic by construction, diffusion cannot generate non-monotonicities as long as the diffusion coefficient remains strictly positive.  
Similarly, the condition of causality imposes an upper bound on the derivative of $\mu(n)$; since the initial configuration satisfies this constraint, the diffusion process cannot produce super-luminal speeds of sound in the diffusion domain as long as 
\begin{equation}
D'' - \frac{D'}{n} \leq 0,
\end{equation}
see App.~\ref{sec:causal_diffusion} for details.

Finally, the diffusion equation conserves the energy density,
\begin{equation}
\varepsilon' - \varepsilon = \int_{n}^{n'} \dd n\, \mu(n),
\end{equation}
and hence preserves thermodynamic consistency within the domain, given that $\varepsilon = -p + \mu n$.  
At the boundaries, however, the energy density may ``flow'' through the edges of the interval,
\begin{align}
    \partial_\tau (\eh - \el) 
    = \partial_\tau \int_{\nl}^{\nh} \! \dd n\, \mu(n,\tau)
    = \left[ D(n)\, \partial_n \mu(n,\tau) \right]_{\nl}^{\nh},
\end{align}
leading to a mild violation of thermodynamic consistency in $\eh$. 
This effect can be mitigated by tapering the diffusion coefficient $D(n)$ to zero near the boundaries or by reinjecting the corresponding energy after the evolution.  
In practice, we find that this deviation alters the energy density at $\muh$ only at the percent level---well below the current theoretical uncertainty of chiral EFT and pQCD calculations at these densities---and can therefore be safely neglected.

\subsubsection{Covariance and correlation length of sound speed}
In order to connect with the existing literature, we highlight the similarities and differences between our constrained GP bridge construction and the standard GP approaches to modeling the speed of sound~\citep{Landry:2018prl, Gorda:2022jvk}. 
In GP-based frameworks, the process is typically defined through a constant or density-dependent correlation length. 
Here, we show that by choosing an appropriate functional form for the diffusion coefficient $D(n)$, one can control the local correlation length of the sound speed while simultaneously maintaining the non-trivial global correlations implied by the underlying physical constraints.

Locally, for slowly varying $D(n)$, the correlation function of the
sound-speed fluctuations, defined as
${\delta c_s^2 \coloneq c_s^2 - \langle c_s^2 \rangle}$, becomes
\begin{align}
\label{eq:covariance}
 &\big\langle \delta c_s^2(n,\tau)\,\delta c_s^2(n',\tau)\big\rangle\\
&= \int d\xi\,d\xi'\,
G(n,\xi;\tau)\,G(n',\xi';\tau)\,
\big\langle \delta c_s^2(\xi,0)\,\delta c_s^2(\xi',0)\big\rangle, \nonumber
\end{align}
where $G(n,\xi;\tau)$ is the Green’s function of the diffusion equation.
For a nearly white initial correlation in the sound speed,
$\langle \delta c_s^2(\xi,0)\,\delta c_s^2(\xi',0)\rangle
\simeq \sigma_s^2\,\delta(\xi-\xi')$,
the evolved correlation function is locally approximated by a Gaussian (see App.~\ref{sec:diffuse_cs2_covariance} for details),
\begin{equation}
\big\langle \delta c_s^2(n,\tau)\,\delta c_s^2(n',\tau)\big\rangle
\propto
\frac{1}{\sqrt{8\pi D(n)\tau}}\,
\exp\!\left[-\frac{(n-n')^2}{8D(n)\tau}\right],
\label{eq:corr}
\end{equation}
with a characteristic correlation length
$\sigma = \sqrt{4D(n)\tau}$.
This provides a direct way to control the local correlation length of
$c_s^2$ through the density dependence of $D(n)$, offering a flexibility
analogous to that of GP kernels,
while simultaneously preserving the global physical constraints
imposed by the diffusion dynamics.

In the following, we aim to reproduce a correlation length that remains logarithmically constant as a function of density in order to both allow for fine structure for the \eos\ at low densities and span the large density range up to pQCD densities; that is, the correlation length from diffusion at all densities corresponds to a fixed fraction of the density, $\sigma/n = c$, with $c$ constant. According to Eq.~\eqref{eq:corr}, such a correlation length is obtained by choosing $D(n)\tau = n^2 c^2 / 4$. In practice, we evolve the diffusion equation from $\tau=0$ to 1, with a density dependent diffusion coefficient $D(n)=n^2c^2/4$.

Fig.~\ref{fig:correlator} shows the two-point correlation function of the sound speed defined in Eq.~\eqref{eq:covariance} evaluated with one of the arguments $n'= 0.8\,\rm{fm}^{-3}$ and as a function of the other density $n$. 
In order to numerically extract the sound speed, we must apply at least a minimal level of diffusion to make the \eos\ differentiable. 
The line corresponding to $c = 1\%$ demonstrates that the self-similar refinement process produces a sharply peaked, yet non-white, correlation function that exhibits a long-range anti-correlation. 
This feature implies that if the \eos\ is stiff at low densities, it tends to become soft at higher densities. 
The reference density in the left figure, $n' \approx 5\,n_s$, is chosen to be representative of the densities expected in a maximum-mass neutron star.
This anticorrelation behavior is consistent with the observation that an \eos\ capable of supporting a two-solar-mass star, which requires stiffening at intermediate densities, must soften at higher densities \citep{Gorda:2022jvk, Komoltsev:2023zor}.

At higher levels of diffusion ($\sigma/n = 10\%, 20\% $), the local correlation approaches a Gaussian form, as expected, and is well reproduced by Eq.~\eqref{eq:corr} (shown as a dotted line). 
As the correlation length increases, the global anticorrelation becomes stronger. 
For sufficiently large correlation lengths ($40\%$), the global structure dominates and the overall correlation exceeds the local Gaussian component that would arise in the absence of global constraints.

The right panel of Fig.~\ref{fig:correlator} shows the correlation function at several reference densities,
$n = [2, 4, 8, 16]\,n_{\rm s} \approx [0.32, 0.64, 1.28, 2.56]\;\mathrm{fm}^{-3}$, for a fixed $\sigma/n = 20\%$. For each density, the dotted line denotes the corresponding Gaussian correlation.

\section{Implementation}
\label{sec:implementation}
Moving to an application of the framework described above, we now specify the low- and high-density inputs and neutron-star observations used in the present work. 

\label{sec:implementation}
\subsection{High- and low-density limits}
\begin{figure}
    \centering
    \includegraphics[width=1\linewidth]{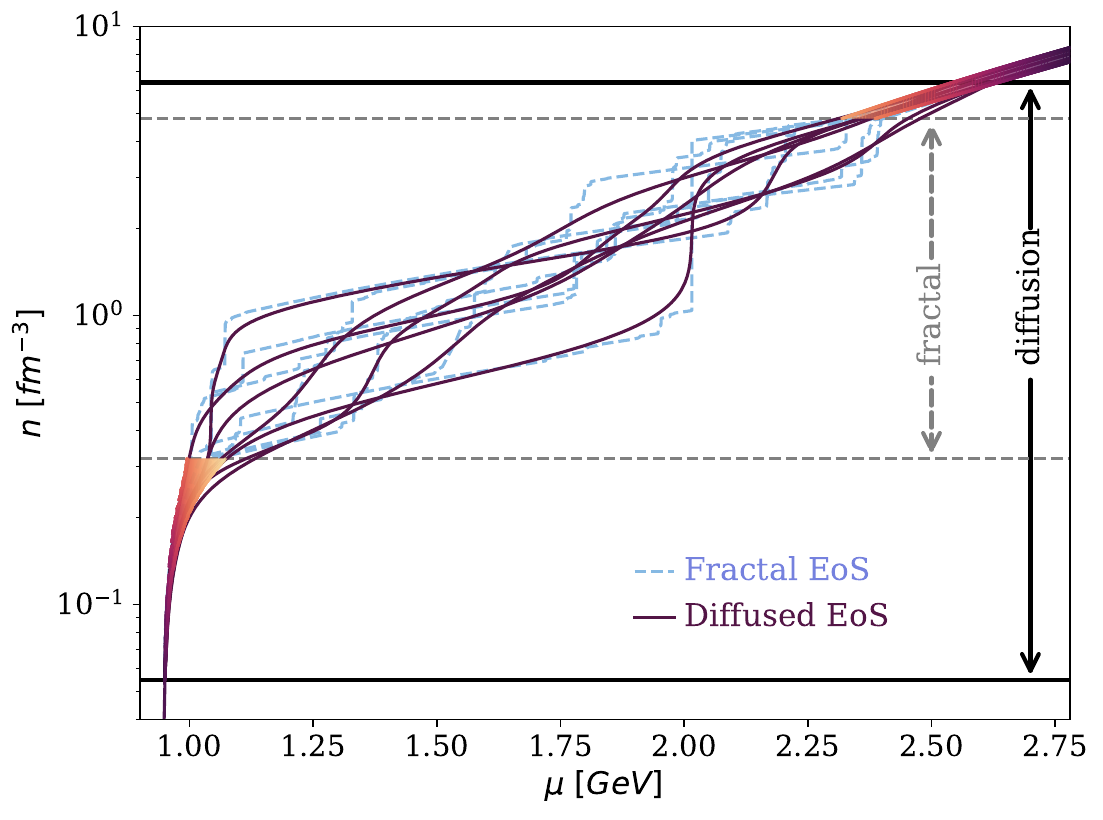}
\caption{The procedure for incorporating the low- and high-density inputs into the construction of the constrained GP bridge. 
The blue dashed lines correspond to the samples of the fractal \eoss\ together with their diffused counterparts represented by purple solid lines. 
The two colored bands correspond to the low- and high-density limits from the chiral EFT and pQCD \eoss. 
The fractal \eos\ is constructed from two endpoints, $\bl$ and $\bh$, obtained from the sampled chiral-EFT and pQCD \eoss\ at the corresponding densities $\nl = 2\,n_{\rm s}$ and $\nh = 30\,n_{\rm s}$. 
The diffusion is then performed over the broader density range $[0.34,\,40]\,n_{\rm s}$ to ensure a smooth transition of $c_s^2$ to both limits.}
    \label{fig:limits}
\end{figure}

The outer crust is modeled using the standard BPS equation of state~\citep{Baym:1971pw}. 
At densities above $n_{\rm atmos} \coloneq 0.34 \,n_s \approx 0.055\ \rm{fm}^{-3}$, we switch to a chiral-EFT-based \eos, describing charge-neutral, $\beta$-equilibrated nuclear matter that includes chiral NN and 3N interactions up to N3LO order \citep{Drischler:2020fvz}. 
In practice, we sample the nuclear \eos\ within the range considered in \cite{Drischler:2020fvz} and extend it up to a density of $\nl \coloneq 2\,n_s \approx 0.32\;\mathrm{fm}^{-3}$, thereby generating a set of nuclear \eoss\ $\mu_{\mathrm{CEFT}}(n)$ over the interval $[n_{\rm atmos},\, \nl]$. 

At high densities we take the  pQCD \eos\ from \cite{Gorda:2021znl} and extend it down to the density $\nh \coloneq 30\,n_{\rm s} \approx 4.8 \ \rm{fm}^{-3}$. 
The pQCD \eos\ has an uncertainty quantified by variation of the renormalization scale, $X$, which we sample from a log-linear distribution~\citep{Gorda:2023usm} to obtain pQCD \eoss\ $\mu_{\mathrm{pQCD}}(n)$. 

The density range between $[\nl, \nh]$ is then populated with the self-similar fractal \eoss.
In the bridge process as described in the previous Section, the primary thermodynamic variables $\mu$, $n$, and $p$ are by construction continuous across the switching points $\nl$ and $\nh$. 
In order to ensure that also the higher derivatives of $\mu(n)$---including the sound speed $c^2_s$---vary smoothly, we extend the range where the diffusion is performed beyond the range of the fractal \eoss.
Specifically, in the range $[n_{\rm atmos},\,\nl]$ we perform diffusion but the initial condition is given by the nuclear \eos. 
Similarly, the initial condition is given by the pQCD \eos\ between $\nh$ and $40 n_\mathrm{s}$.

In Fig.~\ref{fig:limits}, we show several samples of the fractal \eoss\ in the $\mu$--$n$ plane together with their diffused counterparts. The density interval over which the fractal is constructed is indicated by gray dashed lines and corresponds to the range $[\nl,\, \nh]$. 
The black lines show the limits of the density range, $[0.34,\,40]\,n_{\rm s}$, where diffusion is performed.

\begin{figure*}
    \centering
    \includegraphics[width=1.0\linewidth]{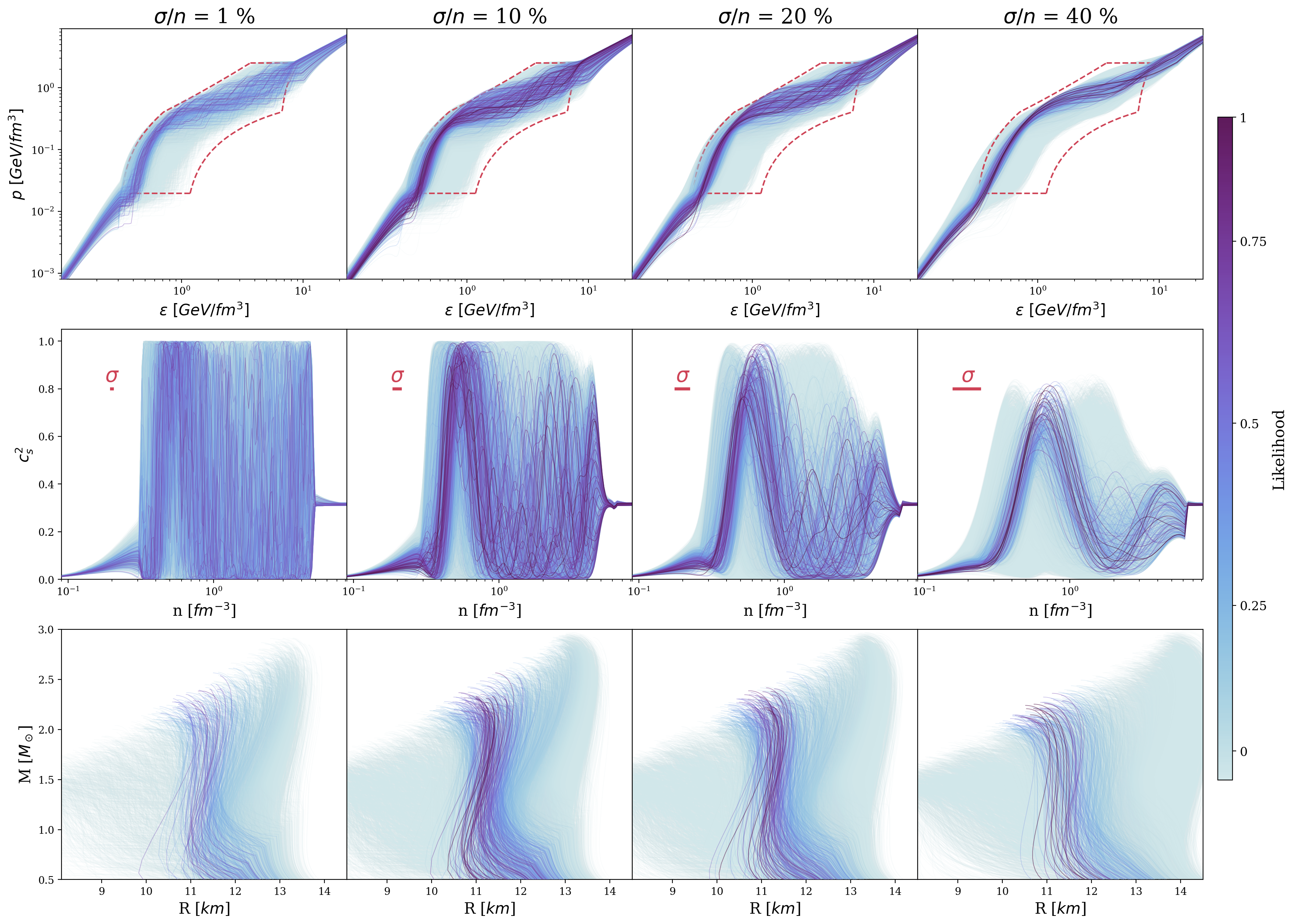}
    \caption{Progression of the posterior equation of state ($\varepsilon$ vs $p$), speed of sound ($c_s^2$ vs $n$) and the mass-radius relation ($M$ vs $R$) as a function of the relative correlation length $\sigma/n = 1\%, 10\%, 20\%$, and $40\%$.  
    Each \eos\ drawn from the prior is colored according to the combined likelihood function. The likelihood function is normalized separately to the maximum likelihood for each fixed correlation length. 
    As the correlation length increases, the global structure of the \eos\ with significant softening at high densities becomes clearer. 
    The bar marked with $\sigma$ indicates the correlation length associated with each column. 
    The relative correlation length corresponds to a fixed length on a logarithmic axis.
    }
    \label{fig:vary_correlation}
\end{figure*}

\subsection{Equation-of-state inference}
To constrain the prior distribution of the bridge process, $P({\rm \eos})$, we condition it on the available neutron-star observations using Bayes’ theorem,
\begin{equation}
P({\rm \eos}\,|\,{\rm Data}) = 
\frac{P({\rm Data}\,|\,{\rm EoS})\,P({\rm \eos})}{P({\rm Data})}.
\end{equation}
For the likelihood, $P({\rm Data}\,|\,{\rm EoS})$, we adopt the likelihood function from~\cite{Gorda:2022jvk} along with some updated measurement data, which incorporates the mass measurements of PSR~J0348$+$0432~\citep{Saffer:2024tlb,Antoniadis:2013pzd} and PSR~J1614$-$2230~\citep{NANOGrav:2017wvv,Fonseca:2016tux,Demorest:2010bx},
the joint mass–radius constraint on PSR~J0740$+$6620 obtained from NICER observations~\citep{Fonseca:2021wxt,Miller:2021qha,Riley:2021pdl, Dittmann:2024mbo}, and the tidal-deformability measurement from GW170817 reported by the LIGO/Virgo collaboration~\citep{LIGOScientific:2018hze}. 
We also incorporate the recent mass measurement (from MeerKAT) and the radius determination (from NICER) for PSR~J0614$-$3329~\citep{Mauviard:2025dmd} as well as the radius determination of  PSR~J0437$-$4715 from NICER~\citep{Choudhury:2024xbk}. 
For \eoss\ with multiple stable branches, we consider only the configurations with largest radius for the same mass, as we are unaware of a production mechanism for the smaller-radius stars.

\section{Results}
\label{sec:results}
In Fig.~\ref{fig:vary_correlation}, we compare the \eos\ of strongly interacting matter from $0.1\, n_s$ to the pQCD regime for four different levels of diffusive correlation, $\sigma/n = 1\%$, $10\%$, $20\%$, and $40\%$.
In the leftmost column, we see that the smallest correlation length covers much of the allowed $\varepsilon$-$p$ region from \cite{Komoltsev:2021jzg} at intermediate densities, while progressively increasing the correlation concentrates the prior more towards the center of the allowed region. 
With increasing correlation length the global structure in $\varepsilon$-$p$ and $c_s^2$-$n$ becomes more clear. 

We also witness a decreased coverage of the full range of $c_s^2$, especially for the highest correlation length of 0.4.
In all cases, the \eoss\ in the posterior distribution exhibit a stiffening immediately beyond the chiral-EFT regime, with shorter correlations showing a larger maximum in $c_s^2$ with a peak at slightly smaller densities.
Beyond this maximum, 
the posterior EoSs show a softer behavior.
In particular, for the two largest correlations considered, the speed of sound exhibits a clear softening onset from $1\, \mathrm{fm^{-3}} \lesssim n \lesssim 2\,\mathrm{fm^{-3}}$ to roughly the pQCD regime at $n \approx 4-6\, \mathrm{fm}^{-3}$, consistent with the behavior observed in earlier works \citep{Annala:2019puf,Gorda:2022jvk,Altiparmak:2022bke,Annala:2023cwx,Komoltsev:2023zor,Finch:2025bao}. 
The shorter correlation lengths that admit for a stronger sound-speed peak at lower densities tend to favor smaller speeds of sound at high densities (best visible for $\sigma/n=20\%$) whereas the larger correlation lengths ($\sigma/n=40\%$) softens the peak, leading to higher speeds of sound at higher densities.
We emphasize that this softening takes place over several correlation lengths in these latter two panels, as illustrated by the line segments representing $\sigma$ in the panels.

We observe that, while the correlation length has a significant effect on the local properties of the \eos, the mass–radius relation---which depends on integrals of the \eos---is largely unaffected by changes in the correlation length. 
However, we do observe that the prior with the largest correlation length reaches slightly larger radii than any of the others, which we attribute to the fact that \eoss\ with longer correlation lengths can remain stiff over an extended density range. 
Finally, the maximum likelihood of a given \eos\ exhibits an inverse dependence on the correlation length (not shown). 
This behavior is expected, as \eoss\ with shorter correlation lengths can rapidly adjust their behavior to remain consistent with a larger fraction of the observational data included in the likelihood, leading to a degree of overfitting.

\begin{figure}
    \centering
    \includegraphics[width=1.0\linewidth]{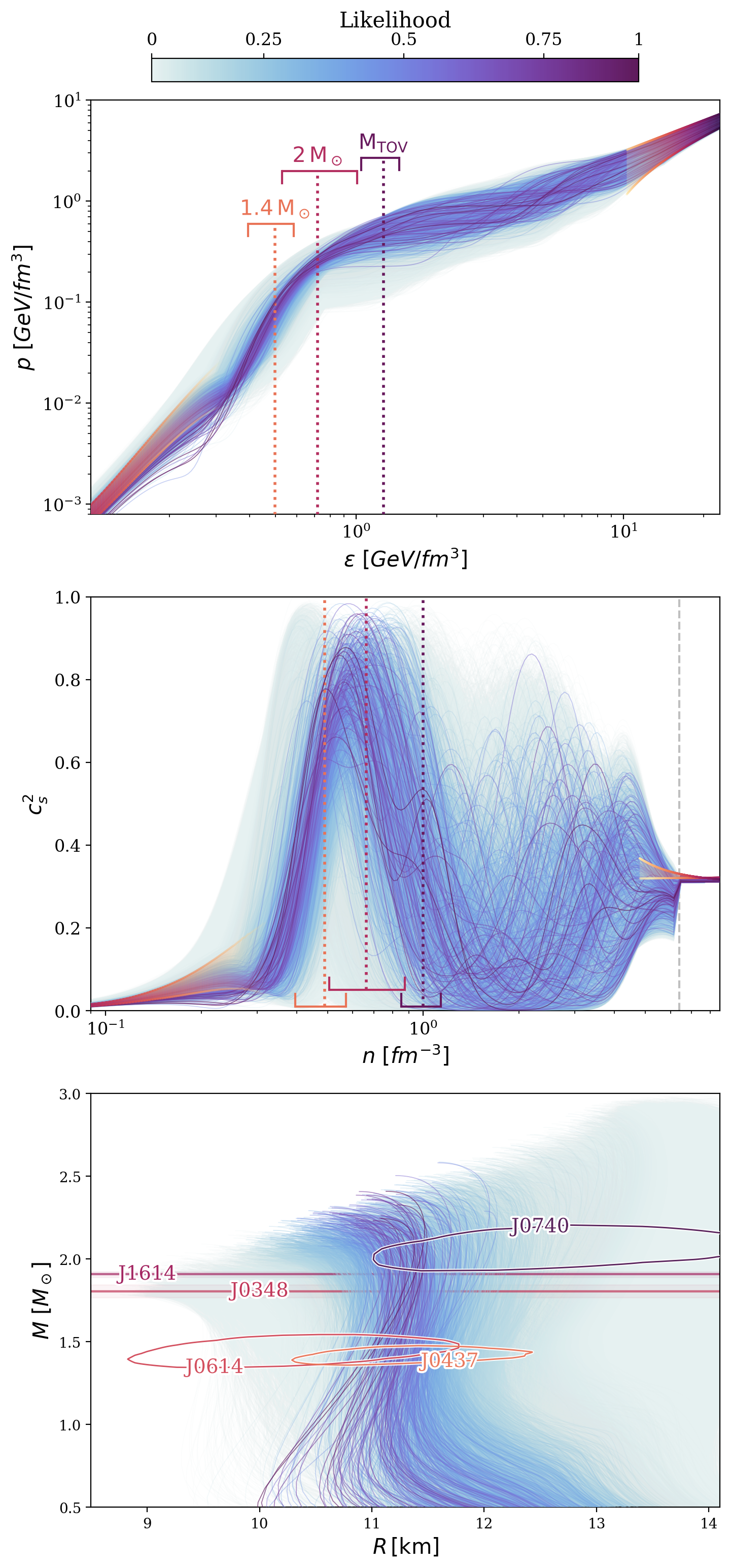}
    \caption{Posterior of $p(\varepsilon)$, $c^2_s(n)$ and $M(R)$ from the hierarchical model discussed in Sec.~\ref{sec:results} with the correlation length taken from a uniform distribution $\sigma/n \in [0.2,0.4].$ The dotted vertical lines and the associated bars correspond to the mean and $1$-$\sigma$ credible intervals for densities reached in the centers of $1.4 M_\odot$, $2 M_\odot$, and $M_{\rm TOV}$ stars, while the contours in the lowest panel represent the 68\% credible regions of the corresponding NICER observations.}
    \label{fig:hierarchical_posterior}
\end{figure}

\begin{figure}
    \centering
    \includegraphics[width=1.0\linewidth]{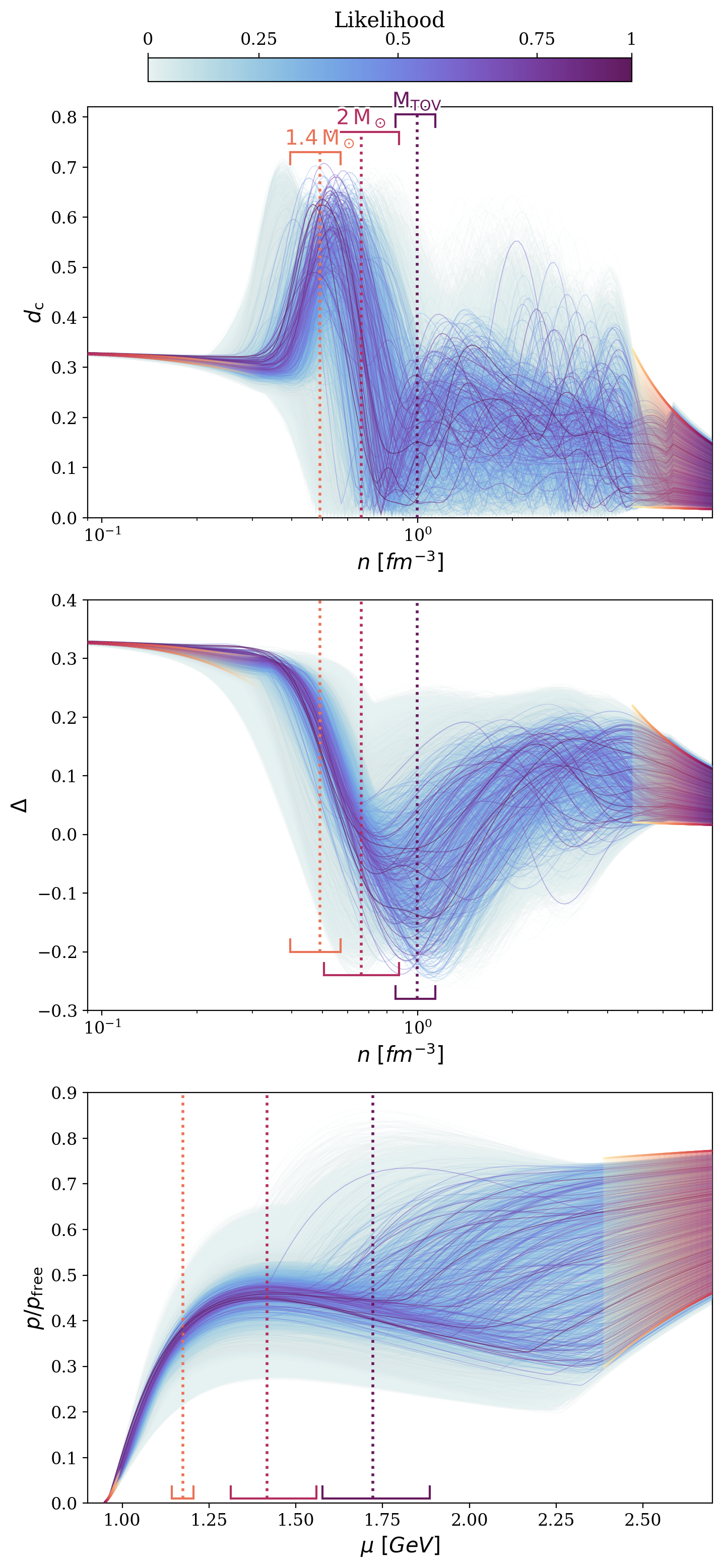}
    \caption{Posterior of $d_\mathrm{c}(n)$, $\Delta(n)$, and $p/p_\mathrm{free}(\mu)$ (see main text for definitions) from the hierarchical model discussed in Sec.~\ref{sec:results} with the correlation length taken from a uniform distribution $\sigma/n \in [0.2,0.4]. $ The dotted vertical lines and the associated bars correspond to the mean and $1$-$\sigma$ credible interval for densities reached in the centers of $1.4 M_\odot$, $2 M_\odot$, and $M_{\rm TOV}$ stars.}
    \label{fig:hierarchical_posterior2}
\end{figure}

\begin{figure*}
    \centering
    \includegraphics[width=1.0\linewidth]{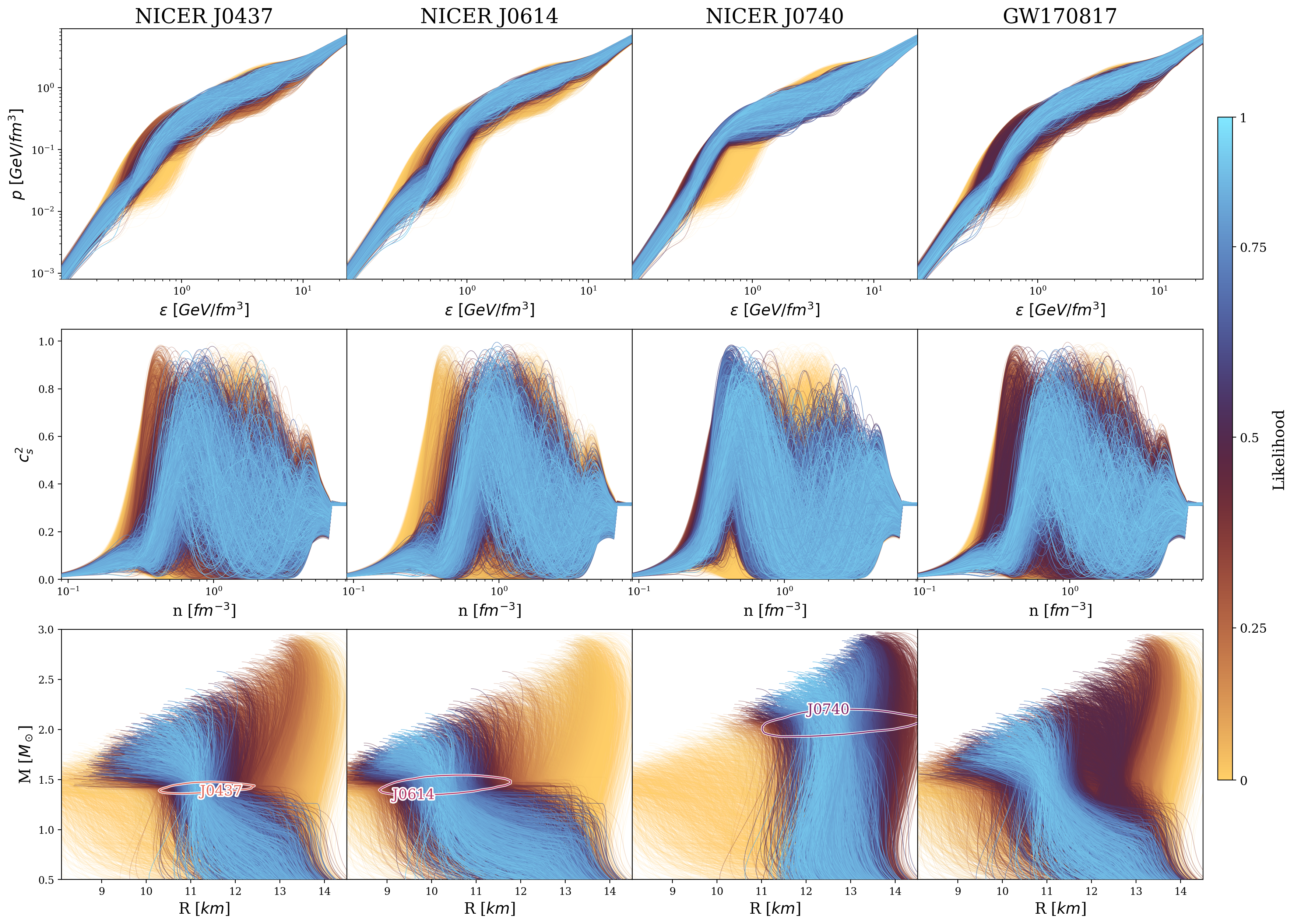}
   \caption{A sample of \eoss, colored according to the likelihood functions derived from various individual X-ray and gravitational-wave observations. NICER J0437 corresponds to the simultaneous mass–radius measurement of PSR J0437$-$4715 from \cite{Choudhury:2024xbk}, NICER J0614 to that of PSR J0614$-$3329 from \cite{Mauviard:2025dmd}, NICER J0740 to that of PSR J0740+6620 from \cite{Dittmann:2024mbo}, and GW170817 to the gravitational-wave observations reported in \cite{LIGOScientific:2018hze}. The prior used here is the hierarchical model discussed in Section~\ref{sec:results}. The contours in the lowest panel represent the 68\% credible regions of the corresponding NICER observations.}
    \label{fig:diff_measurments}
\end{figure*}
In Fig.~\ref{fig:hierarchical_posterior} and Fig.~\ref{fig:hierarchical_posterior2}, we show a posterior sample of 75,000 \eoss\ drawn from a hierarchical model combining different correlation lengths with a uniform range of $\sigma / n \in [0.2, 0.4]$.
We again see a stiffer behavior at lower densities, driven by the high-mass pulsar constraint, followed by a softer average behavior at higher densities. 
In \citet{Fujimoto:2022ohj, Annala:2023cwx}, the trace anomaly  $\Delta \coloneq 1/3 - p/\varepsilon$, the conformal distance $d_\mathrm{c} \coloneq \sqrt{\Delta^2+ (\mathrm{d}\Delta/\mathrm{d}\log\varepsilon)^2}$, and the pressure normalized to that of a free quark gas at the same baryon chemical potential, $p/p_{\rm free}$, were identified as observables sensitive to the phase structure of matter. 
The trace anomaly approaches zero in pQCD, which is expected to be a generic feature of nearly conformal quark matter. 
This behavior is observed at high densities in Fig.~\ref{fig:hierarchical_posterior2}, while at the central densities reached in the most massive stars negative trace anomalies are favored. 
Consistent with \cite{Annala:2023cwx}, the variable $d_\mathrm{c}$, which peaks around densities reached in $2M_\odot$ stars, decreases at larger densities. 
At the highest densities reached in stable neutron stars, the central densities in most \eoss\ satisfy $d_\mathrm{c} < 0.2$, a value quoted as indicative of quark matter. 
At still higher densities, the values plateau (with some fluctuations) as pQCD densities are approached. 
A corresponding behavior is also seen in $p/p_{\rm free}$, which approximately plateaus above chemical potentials of $\mu \sim 1.25\,\mathrm{GeV}$. 
However, given the large uncertainties in the posterior, while the \eos\ may remain approximately constant—as expected for quark matter—a strong first-order phase transition, manifested as a discontinuity in the $\mu$-derivative of $p/p_{\rm free}$, cannot be excluded. 
This is especially true at densities above those reached in the most massive stars, in agreement with~\cite{Komoltsev:2024lcr}.

One advantage of the current technique is that we are able to infer the \eos\ at all densities with a unified prior. 
In particular, we can study how the individual observations constrain the high-density part of the \eos\ beyond the stable neutron-star branch. 
The individual likelihood functions from selected observations are compared in Fig.~\ref{fig:diff_measurments}. 
We see that all the measurements constrain the \eos\ even at densities that go well beyond the densities reached in the corresponding stars, due to the global thermodynamic correlations.
This is particularly visible for PSR~J0614$-$3329, where \eoss\ with the lowest pressures above $\varepsilon \gtrsim 2\, \mathrm{GeV/fm}^{3}$ receive negligible likelihood.
We also find it interesting that all of the measurements besides PSR~J0740$+$6620 resemble each other both qualitatively and quantitatively, favoring smaller radii for the most massive stars, whereas PSR~J0740$+$6620 suggests larger radii. 
The highest likelihood \eoss\ from all three other measurements are consistent among each other but inconsistent with the highest-likelihood \eoss\ from PSR~J0740$+$6620.
It will be interesting to see whether future, more precise, radius and tidal deformability measurements remain consistent with each other.

\section{Conclusions}

\begin{figure*}
    \centering
    \includegraphics[width=0.9\linewidth]{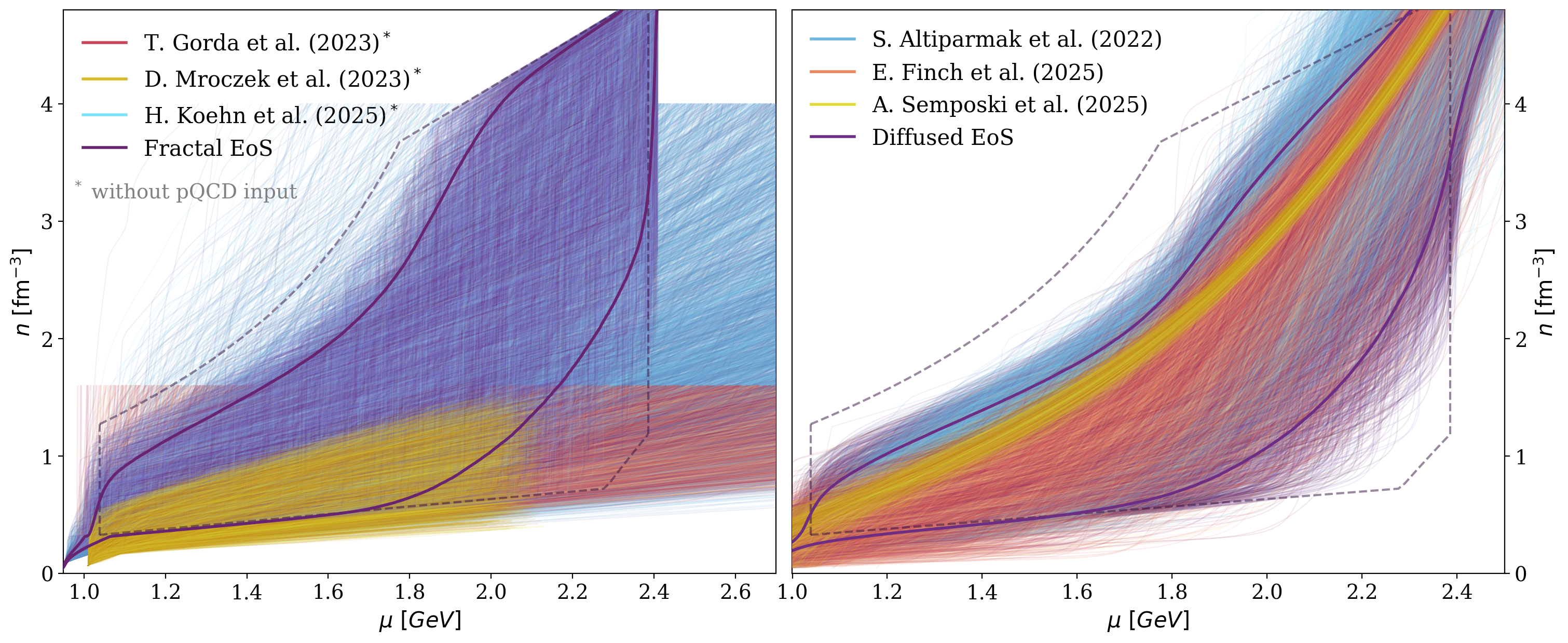}
    \caption{A comparison of the sample of priors from previous works with the current method. 
    (Left) The most generic non-diffused fractal \eos\ prior compared to priors of \cite{Gorda:2021znl}, \citet{Mroczek:2023zxo}, and \cite{Koehn:2024set}. 
    The dashed lines correspond to the boundary of the allowed region when varying the pQCD range of  $X \in [1/2, 2]$ as in \citet{Gorda:2022jvk}. 
    The thick violet line corresponds to the 2-$\sigma$ credible interval for the fractal prior. 
    (Right) The diffused GP bridge compared to other priors that include pQCD constraints. The thick lines correspond to the 2-$\sigma$ credible interval for the diffused prior.}
    \label{fig:priors}
\end{figure*}

In this work, we introduced a new nonparametric prior for neutron-star equation-of-state inference based on a constrained Gaussian-process-bridge construction. 
The method generates equations of state that are thermodynamically consistent, mechanically stable, and causal by construction, while incorporating controlled low- and high-density theoretical inputs from chiral effective field theory and perturbative quantum chromodynamics. 
Unlike existing nonparametric approaches, it does not rely on shooting procedures, intermediate likelihoods, or \emph{ad hoc} switching between \eos\ representations.

A central feature of the construction is its ability to uniformly explore the full physically allowed \eos\ volume between the low- and high-density anchors. 
The resulting prior is maximally flexible within the space of admissible functions, developing structure on all density scales while allowing for controlled local correlations through diffusion. 
This enables a transparent tuning of the sound-speed correlation length and facilitates direct comparison with standard Gaussian-Process-based priors, while retaining the nontrivial global correlations imposed by physical constraints.

These global constraints induce long-range anticorrelations in the speed of sound: equations of state that stiffen at intermediate densities are generically required to soften at higher densities in order to remain consistent with causality, thermodynamic consistency, and the pQCD limit. 
This behavior emerges naturally from the prior and persists even in the limit of vanishing diffusion, demonstrating that it reflects the global structure of the physically allowed equation-of-state space rather than imposed local correlations. 
When conditioned on current astrophysical data, the inferred equations of state must remain sufficiently stiff to support two-solar-mass neutron stars, followed by a systematic softening at higher densities that are still realized within stable neutron stars. 
This qualitative behavior is consistent with previous model-agnostic studies and may be suggestive of phase-transition-like physics in neutron-star cores.

Fig.~\ref{fig:priors} illustrates several qualitative differences between the constrained GP-bridge prior and existing priors in the literature. 
The approaches that do not include the high-density pQCD constraints in the prior level explore a significantly larger function space, large parts of which are excluded; this is exemplified by the comparison of the non-diffused fractal \eos\ with the Gaussian-Process prior of \cite{Gorda:2022jvk} (constructed up to $10\,n_s$) and the piecewise linear $c_s^2$ prior of \cite{Koehn:2024set} (constructed up to $25\,n_s$) in the left panel. 
Also shown in the left panel is the modified Gaussian-Process prior of \citet{Mroczek:2023zxo}, which includes points only on the stable neutron-star branch. The model-informed prior, which is constructed around a rapidly stiffening mean, does not fully explore the space of soft EoSs.
Compared to the approaches that enforce low- and high-density theoretical constraints, the differences are significantly smaller.  
While the differences between the priors arise to some extent from different choices of correlation lengths and different levels of deliberate modeling bias, 
a benefit of our scale-invariant fractal construction is that fills the entire allowed volume without extending beyond it, ensuring that no physically admissible behavior is excluded a priori and enabling a broader exploration of high-density \eos\ trajectories while maintaining a smooth connection to the pQCD anchor. In this sense, the constrained GP-bridge prior should be viewed not merely as a computationally efficient alternative (of the order of seconds for 1000 \eoss), but as a physically motivated framework for encoding prior knowledge about dense matter.  

Looking ahead, the constraining power of this framework is expected to improve significantly as pQCD calculations are extended to higher orders. 
In particular, once pQCD results reach N3LO accuracy with reduced renormalization-scale uncertainties \citep{Gorda:2021znl,Gorda:2023mkk, Karkkainen:2025nkz}, the high-density anchor will become substantially tighter, enabling more precise propagation of first-principles information from quantum chromodynamics to neutron-star densities within the present unified framework.

\section*{Author Contributions}
Authors are listed in alphabetical order.

\begin{acknowledgments}
    We thank  C.\ Drischler, C.\ Ecker, H.~Koehn, D.\ Mroczek,  and A.~Semposki for sharing priors from their respective works. We also thank D.~Furnstahl, T.~Kleppe, and  L.~Rezzolla for interesting discussions. 
    O.K. acknowledges support from the Alexander von Humboldt Foundation through a Humboldt Research Fellowship for Postdoctoral Researchers.
\end{acknowledgments}

\appendix

\section{Expressions for the 3-volume}
\label{sec:3d_volume_eqs}

The volume shown in \cref{fig:3D} (black lines) is defined as follows. For a fixed chemical potential $\mu$, the number density is restricted to the interval $n_{\rm min}(\mu) \le n \le n_{\rm max}(\mu)$. For each allowed pair $(\mu, n)$, the pressure is further constrained to lie between $p_{\rm min}(\mu, n)$ and $p_{\rm max}(\mu, n)$. The analytical expression for $n_{\rm min/max}$ and $p_{\rm min/max}$ are given by:

\begin{equation}
\label{eq:n_min}
n_{\rm min}(\mu)=
\begin{cases}
\nl \, \mu/\mul,
& \mul \le \mu \le \muc, \\[0.6em]
\dfrac{\mu^3 \nh - \mu \muh \left(\muh \nh - 2 \Delta p\right)}
      {\left(\mu^2 - \mul^2\right)\muh},
& \muc < \mu \le \muh .
\end{cases}
\end{equation}

\begin{equation}
\label{eq:n_max}
n_{\rm max}(\mu)=
\begin{cases}
\dfrac{\mu^3 \nl - \mul \mu \left(\mul \nl + 2 \Delta p\right)}
      {\left(\mu^2 - \muh^2\right)\mul},
& \mul \le \mu < \muc , \\[0.6em]
\nh \, \mu/\muh,
& \muc \le \mu \le \muh .
\end{cases}
\end{equation}

\begin{equation}
\label{eq:pmin}
p_{\rm min}(\mu)
=
\pl
+
\frac{\mu^2-\mul^2}{2\mu}
\, n_{\rm min}(\mu).
\end{equation}

\begin{equation}
\label{eq:pmax}
p_{\rm max}(\mu,n)=
\begin{cases}
\pl
+
\dfrac{\mu^2-\mul^2}{2\mu}\, n,
& n \le n_c(\mu), \\[0.8em]
\ph
-
\dfrac{\muh^2-\mu^2}{2\mu}\, n,
& n > n_c(\mu).
\end{cases}
\end{equation}

With the following auxiliary definitions: 
\begin{equation}
\label{eq:muc}
\muc =
\sqrt{
\frac{\mul \muh \left(\muh \nh - \mul \nl - 2 \Delta p\right)}
     {\mul \nh - \muh \nl}
     } ,
\end{equation}
\begin{equation}
\Delta p = \ph - \pl ,
\end{equation}
\begin{equation}
\label{eq:nc}
n_c(\mu)
=
\frac{n_{\rm max}(\mul)}{\mul}\,\mu
=
\frac{n_{\rm min}(\muh)}{\muh}\,\mu .
\end{equation}

\section{Uniform sampling in the allowed 3-volume}
\label{sec:sampling_method}
In order to achieve a uniform sampling on the volume of allowed points of \eos\ $\vlh$, which we denote $\Ulh$, using the transformation method, we perform the following steps:
\begin{enumerate}
    \item Draw $\mu \in [\mul, \muh]$, from the marginalized distribution whose density is given by the area of the triangular slice of $\vlh$ at the value $\mu$; namely, 
    \begin{equation}
        P(\mu) \coloneq \int \dd n \,\dd p \,\, \Ulh
    \end{equation}
    \item Draw $n$ from the conditional marginalized distribution
    \begin{equation}
        P(n | \mu) \coloneq \int \dd p \,P( n, p | \mu ),
    \end{equation}
    \item Finally draw $p$ from $P(p | n, \mu )$, i.e., from uniform distribution between $p_{\rm min}(\mu)$ and $p_{\rm max}(\mu, n)$.
\end{enumerate}
These steps are made simpler by the fact that the slice of $\vlh$ at fixed $\mu$ is a triangle, making it trivial to find $P(p|n,\mu)$ and $P(n|\mu)$. 
Lastly, the marginalized distribution $P(\mu)$ can be straightforwardly computed and has the following form
\begin{equation}
    P(\mu) = 
    \begin{cases}
    \mu \frac{\mu^2 - \mul^2}{\muh^2 - \mu^2} C_1, \quad \mul < \mu < \mu_c, \\
    \mu \frac{\muh^2 - \mu^2}{\mu^2 - \mul^2} C_2, \quad \mu_c < \mu < \muh,
    \end{cases}
\end{equation}
where $C_1$ and $C_2$ are constants ensuring continuity at $\mu_c$. 
For convenience, we normalize the distribution such that $P(\mu_c) = 1$. 
Writing this distribution as a function of $\mu^2$, we have
\begin{equation}
    P(\mu^2)d\mu^2 = d\mu^2
    \begin{cases}
    \left[
    \frac{\mu^2 - \mu_L^2}{\mu_c^2 - \mu_L^2}
    \right] \!\! \left[
    \frac{\mu_H^2 - \mu_c^2} {\mu_H^2 - \mu^2}
    \right]
    , \quad \mu_L^2 < \mu^2 < \mu_c^2, \\
    \left[
    \frac{\mu_H^2 - \mu^2}{\mu_H^2 - \mu_c^2}
    \right]  \!\! 
    \left[
    \frac{\mu_c^2 - \mu_L^2}{\mu^2 - \mu_L^2}  
    \right] ,
    \quad \mu_c^2 < \mu^2 < \mu_H^2.
    \end{cases}
\end{equation}

To sample this distribution, we transform to a variable $X(\mu^2) \coloneq  \int\dd \mu^2 \, P(\mu^2)$. The explicit form of $X(\mu^2)$ can be found analytically
\begin{widetext}
\begin{align}
    X(\mu^2) ={}& 
    \begin{cases}
    \displaystyle
        \frac{\muh^2 - \mu_c^2}{\mu_c^2 - \mul^2}\left[ -(\mu^2 -\mul^2) - (\muh^2 -\mul^2)~\log \left( \frac{\muh^2 - \mu^2}{\muh^2 - \mul^2} \right) \right], \quad \mul^2 \leq \mu^2 \leq \mu_c^2, \\[2ex]
    \displaystyle
        \frac{\mu_c^2 - \mul^2}{\muh^2 - \mu_c^2} \left[ -(\mu^2-\mu_c^2) + (\muh^2-\mul^2) ~ \log \left( \frac{\mu^2-\mul^2}{\mu_c^2 - \mul^2} \right) \right], \quad \mu_c^2 \leq \mu^2 < \muh^2.
    \end{cases}
\end{align}
Finally, in order to sample $P(\mu^2)$ we sample $X$ from a uniform distribution and then transform back to $\mu^2$ with the inverse function
\begin{align}
    \mu^2(X) ={}& 
    \begin{cases}
    \displaystyle
       \muh^2 + (\mu^2_h - \mul^2) ~ W\!\!\left( -\exp\left\{-\frac{X\frac{\mu_c^2-\mul^2}{\muh^2-\mu_c^2}+(\muh^2-\mul^2)}{\muh^2-\mul^2}\right\} \right) , \quad \mul^2 \leq \mu^2 \leq \mu_c^2, \\[2ex]
    \displaystyle
       \mul^2 - (\mu^2_h - \mul^2) ~ W\!\!\left( \frac{\mul^2-\mu_c^2}{\muh^2-\mu_c^2} ~ \exp\left\{ \frac{X\frac{\muh^2-\mu_c^2}{\mu_c^2-\mul^2}+(\mul^2-\mu_c^2)}{\muh^2-\mul^2} \right\} \right), \quad \mu_c^2 < \mu^2 < \muh^2,
    \end{cases}
\end{align}
\end{widetext}
where $W$ is the Lambert-$W$ function.

\section{Condition for the causality of the diffusion}
\label{sec:causal_diffusion}
To ensure that the diffusion process preserves causality, the local sound speed must remain subluminal, $0 < c_s^2 < 1$, if it is initially within this range.  
We therefore recast the diffusion equation in terms of the sound speed $c_s^2(n,\tau)$,
\begin{align}
\partial_\tau c_s^2
&= D\,\partial_{n}^2 c_s^2
+ \left( 2D' - (1 - c_s^2) \frac{2D}{n}\right) \partial_n c_s^2 \label{eq:cs2-collected} \\
&
+ \Big(  \frac{2D}{n^2}-\frac{D'}{n}\Big)(1 - c_s^2)c_s^2 
 + \Big(D'' - \frac{D'}{n} \Big)c_s^2  \nonumber 
\end{align}
We next use a first–touch argument to determine conditions on $D(n)$ that guarantee $c_s^2$ remains within ${0 \leq c_s^2\leq 1}$ for all later times.

If $c_s^2$ has an interior minimum with $c_s^2 = 0$, then $\partial_n c_s^2 = 0$ and $\partial_{n}^2 c_s^2 \ge 0$ there.  
Evaluating Eq.~\eqref{eq:cs2-collected} at that point gives
\begin{equation}
\partial_\tau c_s^2 = D\partial_n^2 c_s^2 > 0,
\end{equation}
which is positive provided that $D(n) > 0$ everywhere in the domain. 
The minimum value of $c_s^2$ can only increase in $\tau$ and therefore $c_s^2$ cannot cross below zero.

Similarly, if $c_s^2$ reaches an interior maximum with $c_s^2 = 1$, then $\partial_n c_s^2 = 0$ and $\partial_{n}^2 c_s^2 \le 0$.  
All terms proportional to $(1 - c_s^2)$ vanish at that point, and Eq.~(\ref{eq:cs2-collected}) reduces to
\begin{align}
\partial_\tau c_s^2
&= D\,\partial_{n}^2 c_s^2
 + \Big(D'' - \frac{D'}{n} \Big)c_s^2 
\label{eq:cs2-maximum}
\end{align}
This is guaranteed to be negative as long as the diffusion coefficient satisfies
$$
D'' - \frac{D'}{n} \leq 0
$$
for all $n$ in the interval. Note, however, that the boundary conditions do not necessarily satisfy $c_2^2 < 1$. In this study, no EoS exhibits such behavior due to the extended diffusion range to chiral EFT and pQCD, where the values of the speed of sound are well below unity, as outlined in Section \ref{sec:implementation}.

\section{Covariance of the speed of sound after diffusion}
\label{sec:diffuse_cs2_covariance}
In this appendix we demonstrate that the diffusion equation leads to
Gaussian correlations for the speed of sound at short distances.  
The assumptions entering this derivation are as follows:
\begin{itemize}
    \item Fluctuations of $\mu(n)$ in the ensemble are small compared to its mean, so that the logarithm in the definition of the speed of sound can be expanded;
    \item The ensemble-averaged chemical potential $\bar\mu(n)$ is a slowly varying function of density;
    \item The diffusion coefficient $D(n)$ can be approximated locally as a constant; and
    \item The short-distance correlations of $c_s^2$ can be approximated as nearly white, even though global structure of the \eos\ may induce long-range correlations. On the contrary, we do not assume anything about the correlations of $\mu$, which are nontrivial due to the constraint structure.
\end{itemize}

We compute the covariance of the sound speed at two densities $n$ and $n'$ after a diffusion time $\tau$:
$$
\big\langle \delta c_s^2(n,\tau)\,\delta c_s^2(n',\tau) \big\rangle .
$$
Assuming small fluctuations around the ensemble mean
$\bar\mu(n,\tau) \coloneq \langle \mu(n,\tau) \rangle$,
we write
\begin{equation}
\mu(n,\tau) = \bar\mu(n,\tau) + \delta\mu(n,\tau).
\end{equation}
The speed of sound can then be expanded as
\begin{align}
c_s^2(n,\tau)
&= n\,\frac{\partial \log(\mu(n,\tau))}{\partial n} \nonumber\\
&\approx
n\,\partial_n \log\bar\mu(n,\tau)
+ n\,\partial_n \!\left(\frac{\delta\mu(n,\tau)}{\bar\mu(n,\tau)}\right).
\end{align}
Because the ensemble mean varies slowly compared to individual realizations, its derivative in the second term induces corrections that are higher order and can be neglected. Hence, to the order we are working we can pull it through the derivative to obtain
\begin{align}
c_s^2(n,\tau)
&\approx
n\,\partial_n \log\bar\mu(n,\tau)
+ \frac{n}{\bar\mu(n,\tau)}\,\partial_n \delta\mu(n,\tau) \nonumber\\
&\approx
\langle c_s^2(n,\tau)\rangle + \alpha\,\partial_n \delta\mu(n,\tau),
\end{align}
where $\alpha \coloneq n_0/\bar\mu(n_0,\tau)$ is treated as locally constant.
In this equation, we used that the ensemble mean of the sound speed $c_s^2(n,\tau)$ can be approximated by the derivative of the ensemble mean of $\mu$  
\begin{equation}
\langle c_s^2(n,\tau)\rangle
= n\,\partial_n \log \bar\mu(n,\tau)
+ \mathcal{O}(\delta\mu^2,\;\delta\mu\,\partial_n\bar\mu).
\end{equation}
The correlation function then reads 
\begin{align}
\big\langle \delta c_s^2(n,\tau)\,\delta c_s^2(n',\tau)\big\rangle
&\approx
\alpha^2\,\partial_n\partial_{n'}
\big\langle \delta\mu(n,\tau)\,\delta\mu(n',\tau)\big\rangle.
\end{align}

Writing the evolved correlation function in terms of the initial condition and the Green function of the diffusion equation,
\begin{align}
&\big\langle \delta\mu(n,\tau)\,\delta\mu(n',\tau)\big\rangle \\
&= \int d\xi\,d\xi'\,
G(n,\xi;\tau)\,G(n',\xi';\tau)\,
\big\langle \delta\mu(\xi,0)\,\delta\mu(\xi',0)\big\rangle, \nonumber
\end{align}
we obtain
\begin{align}
& \big\langle \delta c_s^2(n,\tau)\,\delta c_s^2(n',\tau)\big\rangle \\
&=\alpha^2
\int d\xi\,d\xi'\,
\partial_n G(n,\xi;\tau)\,
\partial_{n'} G(n',\xi';\tau)\,
\big\langle \delta\mu(\xi,0)\,\delta\mu(\xi',0)\big\rangle. \nonumber
\end{align}
Using the self-adjoint property of the diffusion operator, we can
shift the derivatives from $(n,n')$ to $(\xi,\xi')$ and partially integrate the derivatives to act on the initial correlation
\begin{align}
&\big\langle \delta c_s^2(n,\tau)\,\delta c_s^2(n',\tau)\big\rangle\\
&= \alpha^2
\int d\xi\,d\xi'\,
G(n,\xi;\tau)\,G(n',\xi';\tau)\,
\big\langle \partial_\xi \delta\mu(\xi,0)\,\partial_{\xi'} \delta\mu(\xi',0)\big\rangle.
\end{align}
Recognizing that  $\partial_\xi \delta\mu(\xi,0)$ corresponds to
fluctuations of the initial sound speed,
\begin{equation}
\big\langle \delta c_s^2(\xi,0)\,\delta c_s^2(\xi',0)\big\rangle
=\frac{1}{\alpha^2}\,
\big\langle \partial_\xi \delta\mu(\xi,0)\,\partial_{\xi'} \delta\mu(\xi',0)\big\rangle,
\end{equation}
we can express the covariance as
\begin{align}
&\big\langle \delta c_s^2(n,\tau)\,\delta c_s^2(n',\tau)\big\rangle\\
&= \int d\xi\,d\xi'\,
G(n,\xi;\tau)\,G(n',\xi';\tau)\,
\big\langle \delta c_s^2(\xi,0)\,\delta c_s^2(\xi',0)\big\rangle. \nonumber
\end{align}

Assuming a nearly white initial correlation function for the sound speed,
\begin{equation}
\big\langle \delta c_s^2(\xi,0)\,\delta c_s^2(\xi',0)\big\rangle
\approx \sigma_s^2\,\delta(\xi-\xi'), \nonumber
\end{equation}
and using the Gaussian Green function of the diffusion equation
\begin{equation}
G(n,\xi;\tau)
= \frac{1}{\sqrt{4\pi D\tau}}\,
\exp\!\left[-\frac{(n-\xi)^2}{4D\tau}\right],
\end{equation}
the integral yields
\begin{equation}
\big\langle \delta c_s^2(n,\tau)\,\delta c_s^2(n',\tau)\big\rangle
\simeq
\frac{\sigma_s^2}{\sqrt{8\pi D\tau}}\,
\exp\!\left[-\frac{(n-n')^2}{8D\tau}\right].    
\end{equation}
Thus, the diffusion equation transforms initially (nearly) white $c_s^2$ fluctuations
into Gaussian correlations with a characteristic correlation length 
\begin{equation}
\sigma = \sqrt{4D\tau}
\end{equation}

At large separations the above approximations break down, and the global structure of the \eos\ may introduce long-range, non-Gaussian correlations.
Nevertheless, at short distances the Gaussian form provides an excellent local approximation.  
The argument extends to spatially varying $D(n)$ as long as $D(n)$ varies slowly over the correlation length.
\newpage

\bibliographystyle{aasjournalv7}
\bibliography{references}

\end{document}